\documentclass{aa}  
\usepackage{graphicx}
\usepackage{txfonts}
\usepackage{natbib}
\usepackage{multirow}
\usepackage{tikz}
\usetikzlibrary{shapes,arrows}
\usepackage{fixltx2e}
\usepackage{xcolor,colortbl}

\definecolor{LightCyan}{rgb}{0.88,1,1}

\begin{document}

\title{Source finding, parametrization and classification for \\ the extragalactic Effelsberg-Bonn HI Survey}

\author{
L. Flöer\inst{1} 
\and 
B. Winkel\inst{2}
\and 
J. Kerp\inst{1}
}

\institute{
Argelander-Institut für Astronomie, Universität Bonn,
Auf dem Hügel 71, 53121 Bonn\\
\email{lfloeer@astro.uni-bonn.de} \\
\and
Max-Planck-Institut für Radioastronomie, Auf dem Hügel 69, 53121 Bonn
}

\date{}


\abstract
{Source extraction for large-scale \ion{H}{I} surveys currently involves large amounts of manual labor. For data volumes expected from future \ion{H}{I} surveys with upcoming facilities, this approach is not feasible any longer.}
{We describe the implementation of a fully automated source finding, parametrization, and classification pipeline for the Effelsberg-Bonn \ion{H}{I} Survey (EBHIS). With future radio astronomical facilities in mind, we want to explore the feasibility of a completely automated approach to source extraction for large-scale \ion{H}{I} surveys.}
{Source finding is implemented using wavelet denoising methods, which previous studies show to be a powerful tool, especially in the presence of data defects. For parametrization, we automate baseline fitting, mask optimization, and other tasks based on well-established algorithms, currently used interactively. For the classification of candidates, we implement an artificial neural network which is trained on a candidate set comprised of false positives from real data and simulated sources. Using simulated data, we perform a thorough analysis of the algorithms implemented.}
{We compare the results from our simulations to the parametrization accuracy of the \ion{H}{I} Parkes All-Sky Survey (HIPASS) survey. Even though HIPASS is more sensitive than EBHIS in its current state, the parametrization accuracy and classification reliability match or surpass the manual approach used for HIPASS data.}
{}


\maketitle

\section{Introduction}

A major task in conducting large-scale, extragalactic \ion{H}{I} surveys is the separation of the \ion{H}{I} emission line from other, unwanted signal present in the data.
For single-dish surveys, the majority of unwanted signal comes from continuum emission --- ground and celestial --- and radio-frequency interference (RFI).
In most cases, these contributions are far brighter than the signal from the \ion{H}{I} emission line present in the data.
Standard single-dish data reduction removes the bulk of the continuum emission and flags data affected by RFI. Nonetheless, neither of these measures is perfect: In the vicinity of strong continuum sources, the final data product still shows non-flat baselines. Unmitigated RFI can, at first glance, mimic the profile shape of extragalactic \ion{H}{I} emission.

Automated source-finding software usually employs some kind of matched filtering to generate a list of candidate detections from the data. Since the residual baseline and RFI present in the data are highly significant signal, they generate a large number of false candidates. To deal with this situation, source extraction for \ion{H}{I} surveys like the \ion{H}{I} Parkes All-Sky Survey (\mbox{HIPASS}, \citealp{2001MNRAS.322..486B}) and the Arecibo Legacy Fast ALFA Survey (\mbox{ALFALFA}, \citealp{2005AJ....130.2598G}) involves a large amount of manual labour:
Even though these surveys use automatic candidate detection, the parametrization and classification of these candidates in true and false positives is carried out manually.

For the Effelsberg-Bonn \ion{H}{I} Survey (EBHIS, \citealp{2011AN....332..637K}), we develop a completely automatic source finding, parametrization, and classification pipeline. While the data complexity for EBHIS is comparable to HIPASS and makes a manual parametrization possible, we are using EBHIS as a testbed for future \ion{H}{I} surveys. The data volume expected from the Square Kilometre Array (SKA, \citealp{2004NewAR..48..979C}) and its survey-oriented pathfinders, the Australian SKA Pathfinder (ASKAP, \citealp{2008ExA....22..151J}) and the upgraded Westerbork Synthesis Radio Telescope (WSRT) with Apertif \citep{2009wska.confE..70O}, will far surpass any prior survey. A reliable source extraction pipeline is necessary to fully exploit these data sets. Developing a fully automated source extraction pipeline for EBHIS gives a first hint at the challenges coming with future instruments.

This paper is organized as follows: In Sect. \ref{sec:ebhis}, we introduce EBHIS and describe the main characteristics of the extragalactic data. We further describe how we simulate data cubes that serve as the basis of the analysis in this paper. In Sect. \ref{sec:sourcefinding}, we explain how we use 2D-1D wavelet denoising to find relevant signals in EBHIS data and generate candidate sources. Section \ref{sec:parametrization} covers the parametrization of detection candidates. We explain how we optimize masks for the candidates, find a reliable baseline solution, and robustly measure the widths of the line profiles in detail. We demonstrate the accuracy of our parametrization scheme by comparing the measured parameters with the input parameters from the simulated data. In Sect. \ref{sec:classification}, we describe our implementation of an artificial neural network to carry out our classification. We investigate the impact of automatic classification on the completeness and reliability. Section \ref{sec:conclusions} closes the paper with our conclusions.

\section{The Effelsberg-Bonn \ion{H}{I} Survey}
\label{sec:ebhis}

The Effelsberg-Bonn \ion{H}{I} Survey is a northern all-sky \ion{H}{I} survey carried out with the Effelsberg 100-m telescope. It is the first large-scale \ion{H}{I} survey to be conducted with modern backends based on field-programmable gate arrays (FPGA) that allow us to spread 16384 channels over 100\,MHz bandwidth \citep{2006A&A...454L..29K}. This gives EBHIS the required spectral resolution for Galactic \ion{H}{I} science, and the redshift coverage for an extragalactic survey of the local volume. With a single backend setup, EBHIS will be the northern counterpart to both the Galactic All-Sky Survey (GASS, \citealp{2009ApJS..181..398M}, \citealp{2010A&A...521A..17K}) and HIPASS.

Data acquisition and reduction for EBHIS are described in \cite{2010ApJS..188..488W}. For the extragalactic survey, we slightly modify the data reduction process and subsequently bin the spectral axis by a factor of eight. We describe these changes in a forthcoming paper once the data are released to the scientific community.
The final data have a spectral resolution of $10.24\rm\,km\,s^{-1}$ and an average noise level of $23\rm\,mJy\,beam^{-1}$. We compare the parameters of the extragalactic EBHIS to other large-scale \ion{H}{I} surveys in Table \ref{tab:survey_comparison}.

\begin{table*}
\caption{Parameters of selected current large-scale \ion{H}{I} surveys.}
\label{tab:survey_comparison}
\centering
\begin{tabular}{lccc}
\hline
\hline
Parameter & EBHIS & HIPASS & ALFALFA \\
\hline
Coverage & $\delta > -5^\circ$ & $\delta < 25^\circ$ & $ 0^\circ < \delta < 36^\circ$\tablefootmark{a} \\
Survey Area & $22\,424\rm\,deg^2$ & $29\,343\rm\,deg^2$ & $7\,074\rm\,deg^2$\\
Angular Resolution & 10\farcm 8 & 15\farcm 5 & 3\farcm 5\\
Spectral Resolution & $10.24\rm\,km\,s^{-1}$ & $26.4\rm\,km\,s^{-1}$ & $5.4\rm\,km\,s^{-1}$\\
Spectral Coverage & $cz < 18\,000\rm\,km\,s^{-1}$ & $cz < 12\,700\rm\,km\,s^{-1}$ & $cz < 18\,000\rm\,km\,s^{-1}$\\
Noise Level & $23\rm\,mJy\,beam^{-1}$ & $13\rm\,mJy\,beam^{-1}$ & $2.4\rm\,mJy\,beam^{-1}$ \\
Source Density & ---\tablefootmark{b} & $0.2\rm\,deg^{-2}$ & $5.8\rm\,deg^{-2}$\tablefootmark{c} \\
\hline
\end{tabular}
\tablefoot{
\tablefoottext{a}{Restricted to two separate areas between Right Ascension 7\fh 5 to 16\fh 5 and 22\fh 0 to 3\fh 0 hours.}
\tablefoottext{b}{Expected to reach about HIPASS density.}
\tablefoottext{c}{Derived from 40\% of the final survey area.}

}
\tablebib{\citet{2004MNRAS.350.1210Z}; \citet{2005AJ....130.2598G}; \citet{2006MNRAS.371.1855W}; \citet{2011AJ....142..170H}}
\end{table*}

\subsection{Simulated data}

To quantify the performance of the various developed algorithms, we create a set of simulated data cubes that are modeled to match the noise properties of EBHIS data.

At the angular resolution of EBHIS, most sources are unresolved and resolved sources are only nearby, bright galaxies. Sources of high signal-to-noise ratio are easy to parametrize and we show that the developed algorithms are highly accurate for bright sources. The developed source finding and parametrization algorithms also make no assumption about the angular extent of the sources. Therefore, we limit our simulations to the more common case of unresolved galaxies.

Since EBHIS observations are carried out in on-the-fly mode, individual spectra are sampled onto a regular grid using a Gaussian kernel to produce the final data cube \citep{2010ApJS..188..488W}. This leads to correlated noise on the scale of the gridding kernel as opposed to the angular resolution of the telescope. To recreate this noise behavior, we first generate a data cube with uncorrelated, Gaussian noise, subsequently convolve it with the gridding kernel, and renormalize it to the correct amplitude of $23\rm\,mJy\,beam^{-1}$. The noise of adjacent channels is uncorrelated because of the large binning factor used for the data.

We do not simulate baselines or other artifacts in the data. Since baselines and artifacts are caused by various processes, it is difficult to identify and simulate a general case that would not bias our results. We would like to emphasize that the algorithms in the pipeline are developed with robustness in mind. Our source finding scheme (see Sect. \ref{sec:sourcefinding} and \citealt{2012PASA...29..244F}) is especially proven to be robust against typical defects known to degrade single-dish data. Furthermore, we do include artifacts from real EBHIS data when investigating automated classification in Sect. \ref{sec:classification}.

We simulate the line profiles of our sources using the code of \cite{2014arXiv1405.1838S}. Their model creates physically motivated \ion{H}{I} line profiles that can vary in rotation velocity, velocity dispersion, and asymmetry. Using a 2D-Gaussian as the beam model, we add the sources into the simulated noise. 
To avoid blending and simplify our statistical analysis, we choose the spatial source positions to be on a grid.
Although blending does occur in current single-dish surveys, it has little effect on the derived physical quantities like the cosmological \ion{H}{I} mass density, $\Omega_{\ion{H}{I}}$ \citep{2003AJ....125.2842Z}. Due to the increased angular resolution, it is of even less concern for upcoming large-scale \ion{H}{I} surveys. \cite{2012MNRAS.426.3385D} predict that confusion is of little concern for the WALLABY survey \citep{wallabyproposal}, and at most 5\% of sources in the much deeper DINGO UDEEP survey \citep{2009pra..confE..15M} are affected by confusion.
We apply a random subpixel offset to each source to avoid perfect sampling of the line profile. 
This would artificially increase the parametrization accuracy (see Sect. \ref{sec:peak_flux}).
The spectral location is chosen randomly.

We simulate two sets consisting of 120 data cubes with 100 simulated sources each. The simulated sources in both sets are uniformly sampled in rotational velocity between $30\rm\,km\,s^{-1}$ and $600\rm\,km\,s^{-1}$. The first set uniformly covers the total flux range between $1\rm\,Jy\,km\,s^{-1}$ and $30\rm\,Jy\,km\,s^{-1}$, whereas the other data set spans a wider range of total fluxes between $30\rm\,Jy\,km\,s^{-1}$ and $300\rm\,Jy\,km\,s^{-1}$. The first set of sources covers the transition where EBHIS goes from 0\% to 100\% completeness and is used to quantify the pipeline performance near the detection limit of the survey. The second, brighter set of sources serves as a benchmark for the performance in the high signal-to-noise regime. These two sets allow us to investigate the performance of the algorithms on a wide range of sources. By simulating 12\,000 sources in the fainter data set, we can determine the 50\% completeness level with approximately $5\sigma$ confidence (see Sect. \ref{sec:completeness}).

\section{Source finding}
\label{sec:sourcefinding}

\begin{figure}
\includegraphics{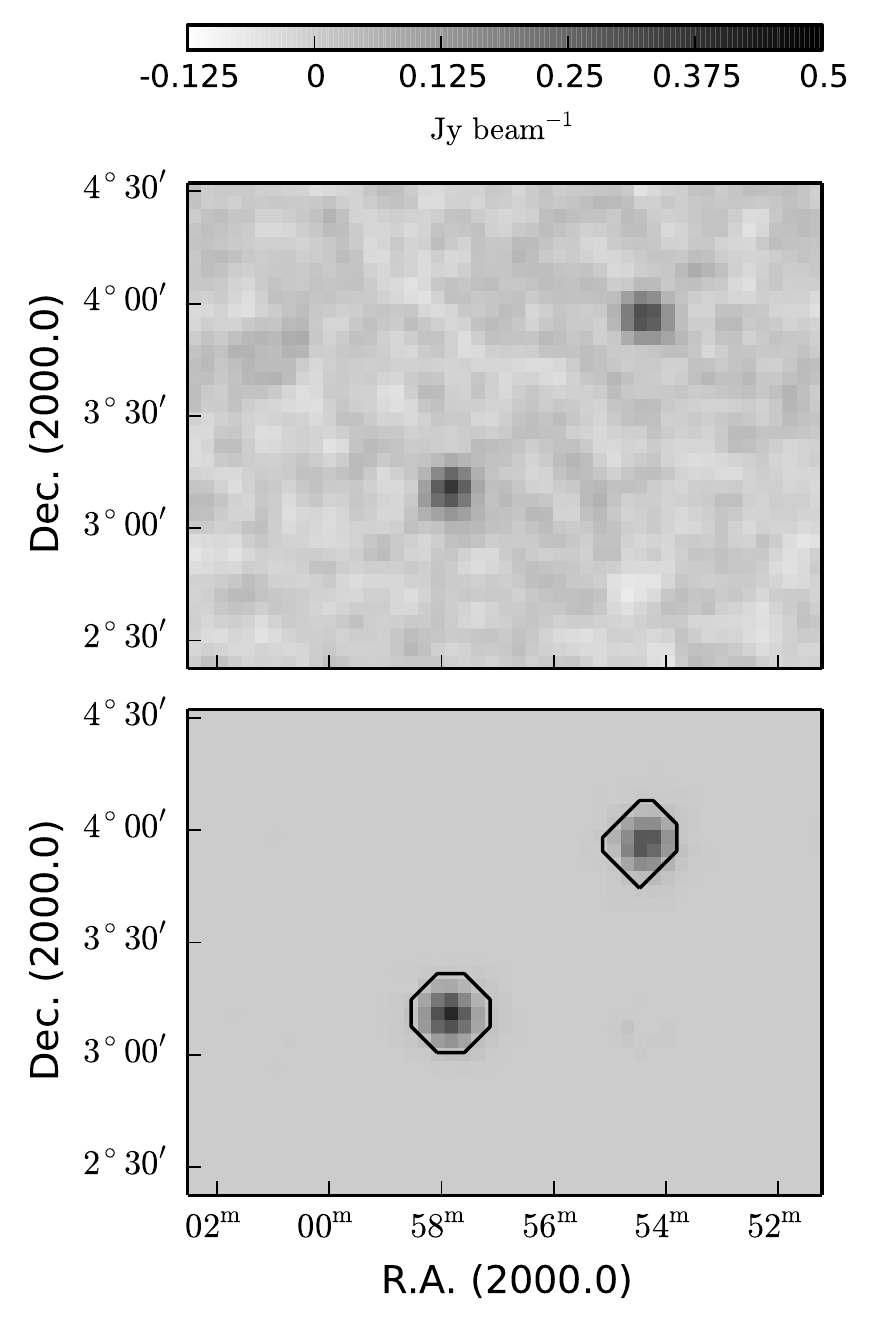}
\caption{Example of a single channel in a simulated data cube from the high signal-to-noise set and its reconstruction. \textit{Top:} Two sources in a simulated data cube. \textit{Bottom:} The same sources in the reconstructed data cube. The contours indicate the initial masks derived from the reconstruction.}
\label{fig:simdata_reconstruction}
\end{figure}

The EBHIS source finding is based on wavelet denoising, which is the removal of noise from the data by means of thresholding insignificant wavelet coefficients and reconstructing the data from only the significant coefficients. A good overview can be found in \cite{5299269}, and details are covered in \cite{Starck:2010uf}.

In \cite{2012PASA...29..244F}, we investigate the performance of a 2D-1D wavelet denoising scheme that is proposed by \cite{2009A&A...504..641S}, which is developed for application to Fermi LAT data \citep{2009ApJ...697.1071A}. This algorithm treats spatial and spectral components separately, respecting the anisotropic nature of \ion{H}{I} line data in the spatial-spectral dimension. This is opposed to 3D denoising, which treats all axes the same. An example of such an implementation can be found in the Duchamp source finding package \citep{2012MNRAS.421.3242W}.

\begin{figure}
\includegraphics{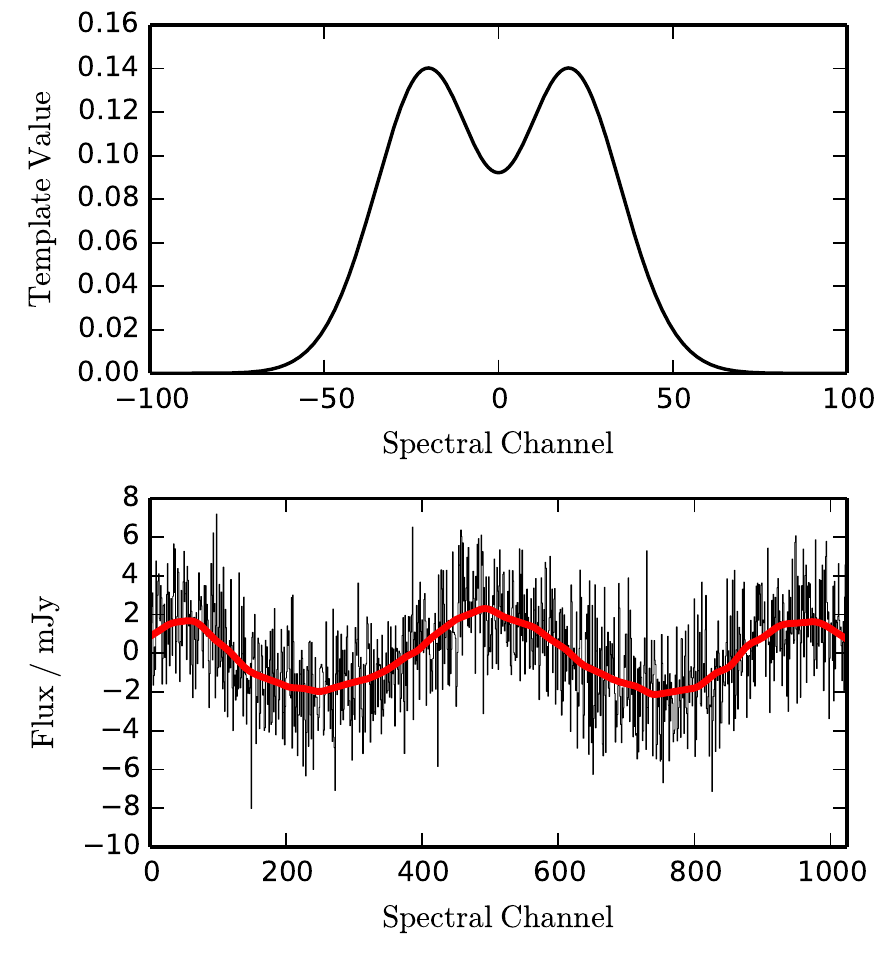}
\caption{Example of the matched filtering process used by \cite{2007AJ....133.2087S} for ALFALFA. \textit{Top:} The matched filtering template used. The template is modeled after the widest profile shown in Fig. 2 of \cite{2007AJ....133.2087S}. \textit{Bottom:} Simulated spectrum with a low frequency ripple with a wavelength of 500 channels. The result of the cross-correlation with the template shown above is shown as the red line.}
\label{fig:matched_filtering}
\end{figure}

Wavelet denoising has further advantages. Single-dish observations often suffer from baseline ripples, which are caused by standing waves between primary focus and apex of the dish. As the amplitude and phase of standing waves depends on many factors, they are difficult to model and remove during the data reduction process. These standing waves represent highly significant signal, which is picked up by simple thresholding of the data.
Sophisticated techniques based on matched filtering are also not resilient against defects in the data. For example, \cite{2007AJ....133.2087S} develop a matched filtering technique. They use templates derived from Hermite polynomials to match the typical shapes of galaxy profiles. We illustrate the issue with matched filtering in Fig.~\ref{fig:matched_filtering}. Since matched filtering is equivalent to a weighted sum, the method does not suppress large-scale ripple in the data.
In wavelet denoising, the data is analyzed at different scales, which can be used to mitigate such effects if the typical scale of the ripple is known. \cite{2012PASA...29..244F} show that the 2D-1D scheme is well-suited to handle common single-dish data defects.

To perform accurate thresholding of the wavelet coefficients, the noise level for a given wavelet scale has to be estimated. As the noise in EBHIS data cubes is spatially correlated on scales of the gridding kernel, determination of the noise level becomes non-trivial. \cite{RSSB:RSSB071} propose to measure the noise-level for a given scale $\sigma_j$ from the wavelet coefficients directly by using the robust MAD\footnote{Median absolute deviation} estimator
\begin{equation}
    \sigma_j = 1.483\,\text{median}\left|w_{j}\right|
\end{equation}
where $\left|w_j\right|$ denotes the absolute value of the wavelet coefficients at scale $j$. This approach only works well for wavelet scales with a large number of independent coefficients and a sparse signal. For increasingly larger scales, adjacent wavelet coefficients become more correlated and in real data, the large-scale signal from the baseline affects virtually every coefficient. This strongly biases the noise level estimation. We therefore perform a wavelet decomposition of an empty, simulated data cube with realistic noise and derive a relation between data noise $\sigma$ and the noise at each wavelet scale: $\sigma_j = \alpha_j\sigma$. This approach allows us to generate sufficiently large data cubes to measure the noise in all relevant scales with high accuracy. Since the coefficients $\alpha_j$ depend only on the type of transform and wavelet chosen, this noise modeling has to be performed only once.

We decompose the data using the 2D-1D decomposition with four spatial and seven spectral scales. This scale selection covers all relevant scales in our data that contain galaxy signals and suppresses large-scale fluctuations. To minimize the impact of the ringing phenomenon in thresholding of un-decimated wavelet transforms \citep{starck2007undecimated}, we reconstruct the data iteratively. We start with a very high threshold at a value, such as, $50\sigma_j$, and lower it in subsequent iterations. To extract as much signal as possible from the data, we perform multiple iterations at the lowest threshold. Additionally, we enforce a positivity on the solution by setting all negative values in the reconstruction to zero  after each iteration. This improves the quality of the denoising process significantly.

Once the data is reconstructed, we use an object generation code developed by \cite{2012PASA...29..251J}.
Each cluster of significant voxels in the data cube is connected to form a source candidate. The code also performs a simple size-thresholding to reject candidates, which either occupy only a single spectral channel or are detected in less than six spatial pixels. The source finding process often misses the faint emission connecting the two edges of very wide, low signal-to-noise profiles. Broken-up sources are joined during the object generation stage: If there are two candidates at the same spatial location but separated by less than eleven spectral channels, they are connected to form a single candidate. Figure~\ref{fig:simdata_reconstruction} shows an example of simulated data and reconstruction.

\section{Parametrization}
\label{sec:parametrization}

\begin{figure}
\centering

\tikzstyle{decision} = [diamond, draw, 
    text width=4.5em, text badly centered, node distance=2cm, inner sep=5pt]
\tikzstyle{block} = [rectangle, draw, 
    text width=10em, text centered, rounded corners, minimum height=2em, fill=white]
\tikzstyle{line} = [draw, -latex']

\begin{tikzpicture}[node distance = 1cm, auto]
    \node [block] (genspec) {Extract spectrum};
    \node [block, below of=genspec] (estbl) {Estimate baseline};
    \node [block, below of=estbl] (optzm) {Optimize spectral mask};
    \node [block, below of=optzm] (genmom) {Generate map};
    \node [block, below of=genmom] (fitcen) {Fit centroid and shape};
    \node [block, below of=fitcen] (optxym) {Optimize spatial mask};
    \node [decision, below of=optxym] (dex) {Extended?};
    \node [block, below of=dex, node distance=2cm] (mfin) {Measure parameters};
    \path [line] (genspec) -- (estbl);
    \path [line] (estbl) -- (optzm);
    \path [line] (optzm) |- +(3,0)  |- node [near start, anchor=west] {$2\times$} (estbl);
    \path [line] (optzm) -- (genmom);
    \path [line] (genmom) -- (fitcen);
    \path [line] (fitcen) -- (optxym);
    \path [line] (optxym) -- (dex);
    \path [line] (dex) -- +(4,0) -- node [anchor=west] {$2\times$} +(4,7) -- (genspec); 
    \path [line] (dex) -- (mfin);
\end{tikzpicture}

\caption{Flow diagram of the parametrization process for each source candidate.}
\label{fig:parflow}

\end{figure}
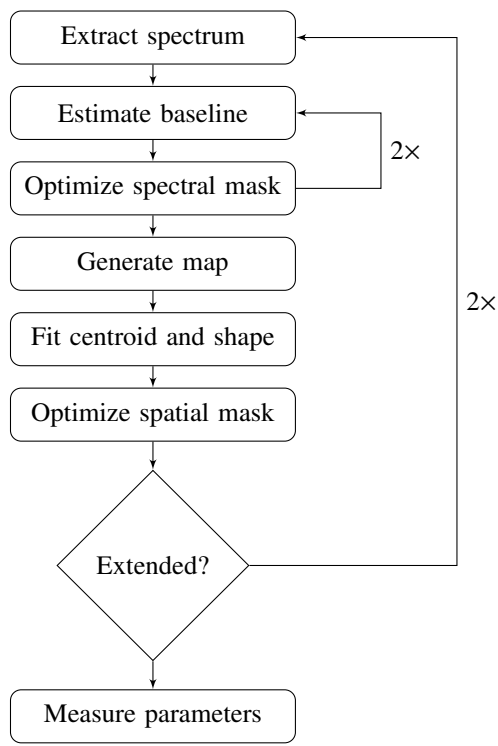

For each region identified as a potential source, we run a set of parametrization steps. We give a brief overview of the pipeline and describe the various algorithms in more detail below. The general implementation of the pipeline is shown in Fig. \ref{fig:parflow}. The first part of the pipeline is concerned with finding optimal masks for the emission. Since the individual mask and centroid optimization steps depend on each other we perform multiple iterations of the critical steps.

The pipeline first extracts a spectrum at the peak of the spatial brightness distribution (peak spectrum) for each candidate and estimates a baseline using smoothing splines (Sect. \ref{sec:baseline_estimation}). Using this spectrum, the pipeline optimizes the spectral mask of the candidate by measuring the width of the profile (Sect. \ref{sec:linewidth_measurement}). Using this optimized spectral mask, the pipeline generates a velocity integrated map of the candidate and fits the centroid and shape of the source (Sect. \ref{sec:centroid_fitting}). Afterwards, the spatial mask is optimized. Before the second outer iteration, the pipeline uses the fit to the shape of the candidate to decide whether the candidate is be treated as unresolved or resolved (Sect.~\ref{sec:centroid_fitting}). In the latter case, the second outer iteration is performed using an integrated spectrum instead of a peak spectrum. The integrated spectrum is generated by summing the flux inside the spatial mask in each spectra channel. This is important to capture the full spectral extent of the source during the spectral mask optimization. After the mask optimization, the final parameters are measured.

\subsection{Baseline estimation}
\label{sec:baseline_estimation}

Spectroscopic surveys require baseline fitting to separate the line of interest from the continuum emission in the data. This is often achieved by masking the spectral range that contains significant emission and by modeling the remaining data with a polynomial. For EBHIS, the majority of the continuum emission, which is ground and celestial, is removed during the standard data reduction process. Still, especially in the vicinity of bright continuum sources, the spectra often do not have a flat baseline and, therefore, require additional treatment.

When done interactively, the user can choose the appropriate degree of the polynomial which characterizes the baseline best but does not overfit the data. To automate the process of polynomial fitting we fit a set of polynomials with increasing degree to the data. The best fitting polynomial is chosen by using the corrected version of the Akaike Information Criterion \citep[${\rm AIC}_c$,][]{1100705, hurvich1989regression}. In the case of least-squares fitting, the ${\rm AIC}_c$ can be calculated from the $\chi^2$ of the fit, the number of free parameters $k$, and the number of data points $N$ by
\begin{equation}
{\rm AIC}_c = N \ln\left(\chi^2\right) + 2k + \frac{2k(k+1)}{N-k-1}\quad.
\end{equation}
Here, the first two terms are the classical AIC and the last term is the correction proposed by \cite{hurvich1989regression}. The $\rm AIC_c$ does not provide an absolute measure of goodness of fit. But among a given set of models, the best fitting model has the smallest $\rm AIC_c$. Since polynomials are known to be a good model for baselines, this shortcoming is not problematic. Using the $\rm AIC_c$ for baseline estimation is very fast and works well for sufficiently smooth baselines. Oscillating baselines in the vicinity of strong continuum sources still pose a problem for polynomials: complex signals require very high polynomial degrees, which can make fitting unstable.

As an alternative, we investigate smoothing splines since they have long been used to smooth noisy data \citep{Reinsch:1967fk, reinsch1971smoothing}. Their only free parameter, the smoothing factor, can be determined from the data using the method of generalized cross-validation \citep{craven1978smoothing}. \cite{garcia2010robust} introduces a version of splines that are robust to outliers in the data. They are therefore very well adapted to the realities of automated baseline estimation. The only drawback as compared to polynomial fitting is the increased computational complexity.

\begin{figure}[t]
\includegraphics{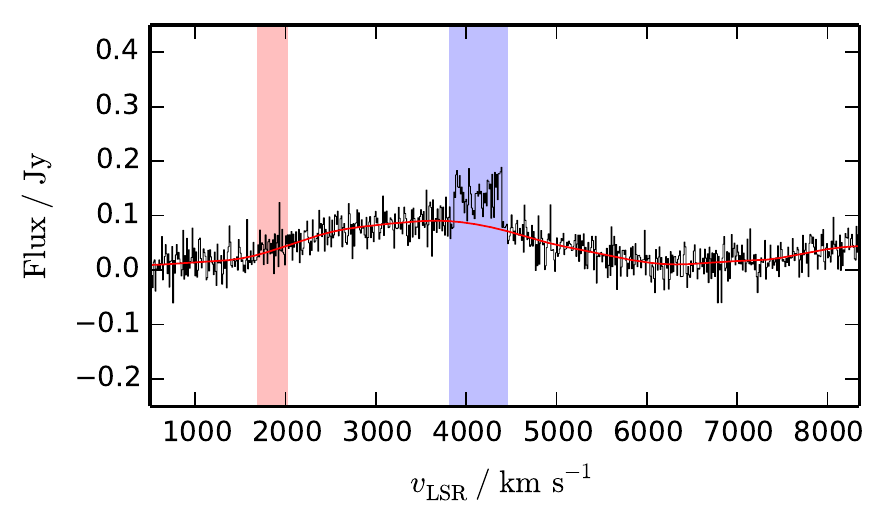}
\caption{Example of a simulated EBHIS spectrum. The blue and red regions indicate the areas where the spectrum has been blanked prior to baseline estimation (see text). The red line shows the baseline solution obtained by spline smoothing.}
\label{fig:baseline_testing_example}
\end{figure}

To assess the performance of the different baseline estimation algorithms, we create a set of 10\,000 artificial EBHIS spectra. Each spectrum contains a simulated galaxy profile with a total flux of $30\rm\ Jy\,km\,s^{-1}$ and a random linewidth between $30\rm\ km\,s^{-1}$ and $600\rm\ km\,s^{-1}$. We then create masks that completely cover the emission and mask an additional, random block of 10 to 20 channels in the spectrum to simulate the impact of missing data, for example, from blanking the emission from the Milky Way. Since we assume the location of the source to be known the actual magnitude of the total flux is not relevant. We choose to keep the total flux fixed to be able to compare the error across different line widths. We generate simulated baselines by adding two $\cos^2$ terms with random phase and frequency. The amplitude has been fixed to twice the noise level of the simulated spectra. This generates a wide variety of baselines without resorting to polynomials which would give an advantage to baseline estimation using polynomials. An example of a simulated spectrum can be seen in Fig. \ref{fig:baseline_testing_example}.

\begin{figure}[t]
\includegraphics{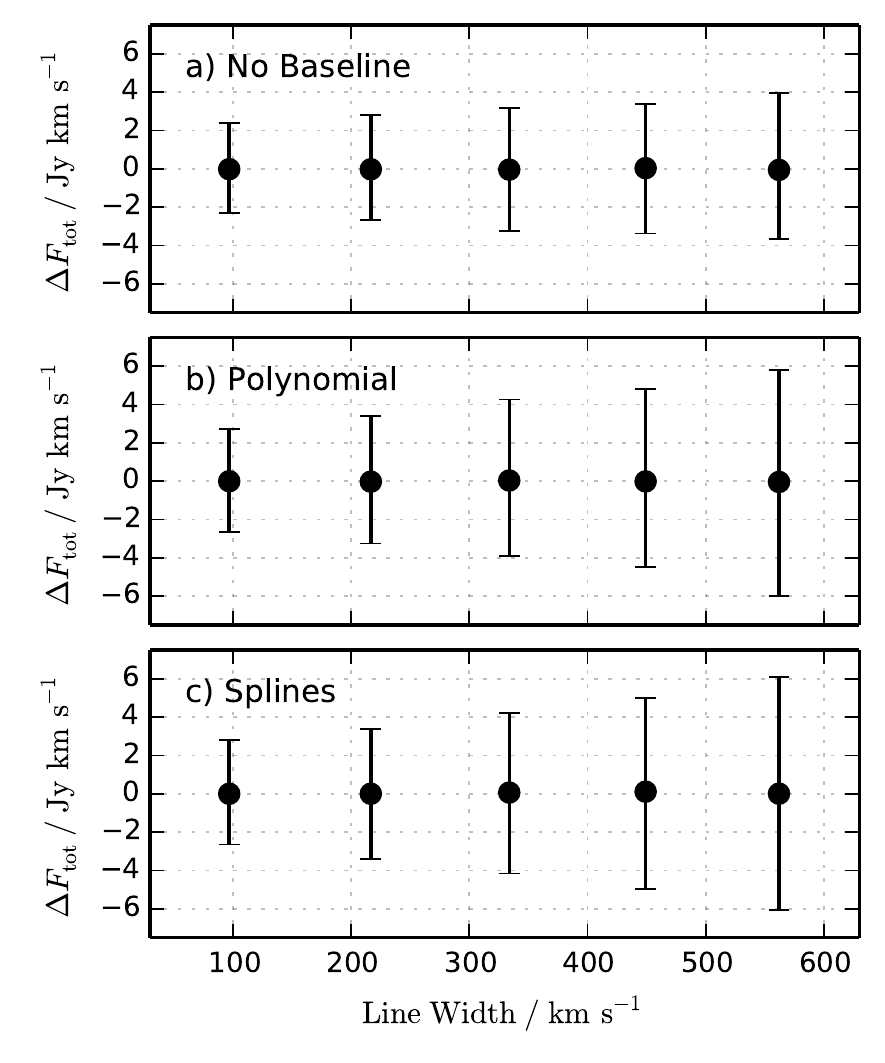}
\caption{Total flux scatter for the investigated baseline fitting procedures. In all panels, the error bars indicate the 95\% confidence region. \textit{Top panel}: Flux measured from a baseline-free spectrum as reference. \textit{Middle panel}: Flux measured from the spectra with polynomial baseline estimation. \textit{Bottom panel}: Flux measured from the spectra with robust smoothing spline estimation.}
\label{fig:baseline_testing_results}
\end{figure}

We summarize our results in Fig. \ref{fig:baseline_testing_results}. To have a benchmark for the performance of the baseline estimation, we first measure the flux from the sources without adding a baseline. Any scatter in the measured flux only comes from the noise in the spectrum. The top panel clearly shows the expected behavior: The flux gets more uncertain for increasing linewidths. We then estimate the baseline using both the polynomial and spline algorithm and measure the flux again. In both cases, the scatter increases as the baseline below the sources has to be extrapolated. From our simulated data, there is no significant difference in the performance of the two algorithms.

When applying the polynomial baseline estimation with $\rm AIC_c$ to real data we notice, that the $\rm AIC_c$ prefers overly complex models for some cases of very flat baselines. This leads to unconstrained solutions, especially in the masked region of the spectrum. Additionally, the polynomial baselines are not robust against outliers in the data. Because of the insignificant difference in performance and the increased stability, we select the robust smoothing splines for our pipeline.

\subsection{Linewidth measurement}
\label{sec:linewidth_measurement}

\begin{figure}[t]
\includegraphics{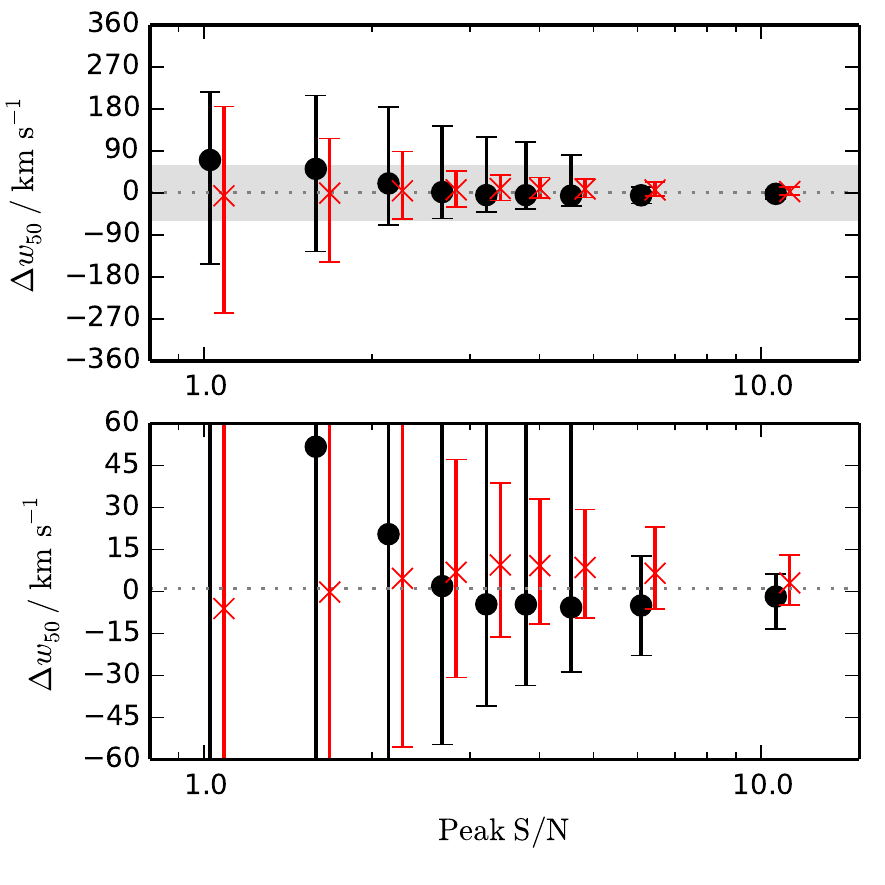}
\caption{Median and 95\% confidence regions for the error in $w_{50}$ for both the classical method (black circles) and our bilateral filtering approach (red crosses) as a function of peak signal-to-noise ratio. Both panels show the same data but at different zoom scales. The shaded region in the top panel indicates the range in $\Delta w_{\rm 50}$ shown in the bottom panel. The red crosses have been offset slightly for clarity.}
\label{fig:linewidth_testing_results}
\end{figure}

The most common way to quantify the width of an \ion{H}{I} profile is to measure its extent at 50\% or 20\% of the peak flux of the spectral profile. If this width is measured at the points where the flux density first rises above the threshold, we speak of a linewidth maximizing algorithm. If the width is measured between the points where the flux density first falls below the threshold, the algorithm is called width minimizing. As the name suggests, the latter algorithm has a tendency to underestimate the profile width and is therefore not well suited for an automated pipeline.

Apart from the peak measurements, there are other methods published in the literature. \cite{2009AJ....138.1938C} measure the linewidth at the points where the flux density first rises above 50\% or 20\% of the mean flux across the profile. This approach has the drawback that an accurate mask for the source has to be known beforehand as the mean flux needs to be measured as well. \cite{0067-0049-160-1-149} fit polynomials to the rising and falling side of the profile and measure the profile width using the fits instead of the actual data to reduce the impact of noise on the measurement. In all aforementioned cases, the estimation of the linewidth has always been interactive. Especially for cases in which the peak signal-to-noise ratio falls below 6 to 7, these linewidth measurements start to become wildly inaccurate as shown by \citet{1986AJ.....91..705B}.

For this reason, it is common to smooth spectra that do not fulfill a certain signal-to-noise ratio. Here, either a fixed kernel or a small set of kernels of varying width are used. Smoothing has the drawback that is smears out sharp edges and thereby lowers the resolution of the line profiles. Instead of using linear filters, we use a bilateral filter (\citealp{tomasi1998bilateral}, hereafter TM98). Bilateral filters combine domain and range filtering, in that they not only consider how close a set of values is but also how similar they are in amplitude. This property leaves sharp edges untouched but smooths areas that are similar in amplitude. We follow the example in TM98 and choose Gaussian kernels for both domain and range filtering. We fix the dispersion for the domain filter to $\sigma_d = 3$ channels and adjust the range filter according to the noise level in the spectra. Using simulations, we determine the optimal value to be $\sigma_r = 3\sqrt{2}\sigma_{\rm noise}$. The factor of $\sqrt{2}$ is included since the similarity in amplitude of two values is measured from the difference between them. The difference of two independent Gaussian random variates with equal standard deviation is again a Gaussian random variate with a factor of $\sqrt{2}$ larger standard deviation.

To verify the performance of our approach, we generate 10\,000 synthetic spectra with flat baselines and $23\rm\,mJy$ noise. Using the noise-free profile, we generate masks for the emission and enlarge them by five channels on each side to simulate uncertainty about the true spectral extent of the source. For both the raw and smoothed spectrum, we use a width maximization algorithm and search inwards, starting five channels outside of the mask. Figure~ \ref{fig:linewidth_testing_results} summarizes our findings. The filtered approach is clearly favorable and gives satisfying results down to a peak signal-to-noise ratio of three. We also observe a slight underestimation of the linewidth in the raw spectrum, which is replaced by a slight overestimation from the smoothed spectrum. This bias can be easily corrected, and the 95\% confidence regions obtained are more symmetric and much less biased for a low peak signal-to-noise ratio.

Recently, multiple methods that determine the parameters of the spectral profile by modeling are developed \citep{2014MNRAS.438.1176W, 2014arXiv1405.1838S}. Provided with sensible starting parameters, these methods are suitable for unsupervised parametrization. They are under consideration for inclusion in the final EBHIS pipeline. Our non-parametric estimates could serve as the starting parameters of the fitting process.

\subsection{Centroid and shape}
\label{sec:centroid_fitting}

We determine the angular position and shape from the velocity-integrated map of each source by using two different elliptical Gaussian fits.

First, we measure accurate positions of the candidates by using an iteratively re-weighted elliptical Gaussian fit. In each iteration step, we weigh the data by the fit from the previous step. The iteration is stopped once the fit parameters change less than 1\% between iterations. Using this approach, we avoid a shifting of the centroid coordinates due to extended emission or artifacts near the candidate. During this fit, all parameters --- amplitude, center, major and minor axis, and position angle --- of the elliptical Gaussian are free.

Since the iterative fit contracts around the brightest part of the emission, it does not reflect the shape and orientation very well. To better quantify the extent and orientation of the emission, we perform an additional Gaussian fit. For this fit, we fix the center coordinates to the ones determined from the iterative fit. This second Gaussian fit is performed without weighting. It is therefore more sensitive to the full extent of the emission and better reflects the shape of the source.

Since unresolved sources should have an angular extent compatible with the resolution of the data, we use this second Gaussian fit to determine whether a source is resolved or unresolved. If the major axis is larger than $1.5$ times the angular resolution of the data, the source is treated as resolved during the second iteration of the pipeline.

Although our simulation only includes unresolved sources, preliminary testing on real data has shown that this criterion performs well. Nonetheless, for faint sources, the automatic estimation of the extent can yield wrong results, for example, an unresolved source is treated as being resolved due to a bad fit. This introduces larger parametrization error for faint sources. We therefore include this automatic decision process to obtain more realistic error estimates for faint, unresolved sources.

\subsection{Mask optimization}
\label{sec:mask_optimization}

Regardless of the method used, every source finder includes thresholding of some kind. The masks provided by the source finding algorithms are therefore systematically too small. This can lead to a significant underestimation of the total flux. While testing the Duchamp source finding package, \cite{2012PASA...29..276W} notice a systematic underestimation of the measured total flux. This effect is especially pronounced in faint sources. Since these sources make up the bulk of detected galaxies, any statistical analysis is highly biased.

We alleviate this issue by mask optimization. Using the position determined from the iterative Gaussian fit, we measure the total flux in the velocity-integrated map in increasing radii. Once the flux inside the mask does not increase any more, we stop the process and take this aperture as our new mask. In case of extended sources, we switch to optimizing elliptical apertures, as these describe the typical shape of a galaxy more accurately.

It should be noted that only the very nearby galaxies are expected to be resolved for EBHIS, and the majority will be only marginally resolved or completely unresolved. Even though future surveys with ASKAP and WSRT/Apertif will detect vastly more resolved sources, the fraction of unresolved or barely resolved sources will be similarly high \citep{2012MNRAS.426.3385D}.

\subsection{Final parameters}

After the optimization processes are finished, the final parameters for each candidate are measured. From both the integrated and peak spectrum, we measure the linewidth and total flux. The redshift is measured from the midpoint between the two velocities where the spectrum rises above 50\% of the peak flux. Additionally, we measure various shape parameters of both the line profile and the velocity-integrated map. These are used for classifying the candidate after the parametrization is finished. For the line profile, we measure the skewness and kurtosis. From the velocity-integrated map, we measure the concentration parameter of the source by comparing the solid angle containing 50\% and 80\% of the total flux. We also derive cumulative and differential surface brightness profiles. These profiles are made size-independent by determining the surface brightness at multiples of the semi-major axis of the source. The additional parameters determined from the velocity-integrated map are not expected to contain relevant physical information for unresolved sources. They are nonetheless important for the classification process, as the classification pipeline requires that every candidate is parametrized in the same way. This becomes important in Sect.~\ref{sec:classification}, where we add parametrized artifacts from real EBHIS data. As artifacts typically do show spatial structure, these parameters have strong discriminatory power.

\begin{figure*}[t]
\includegraphics{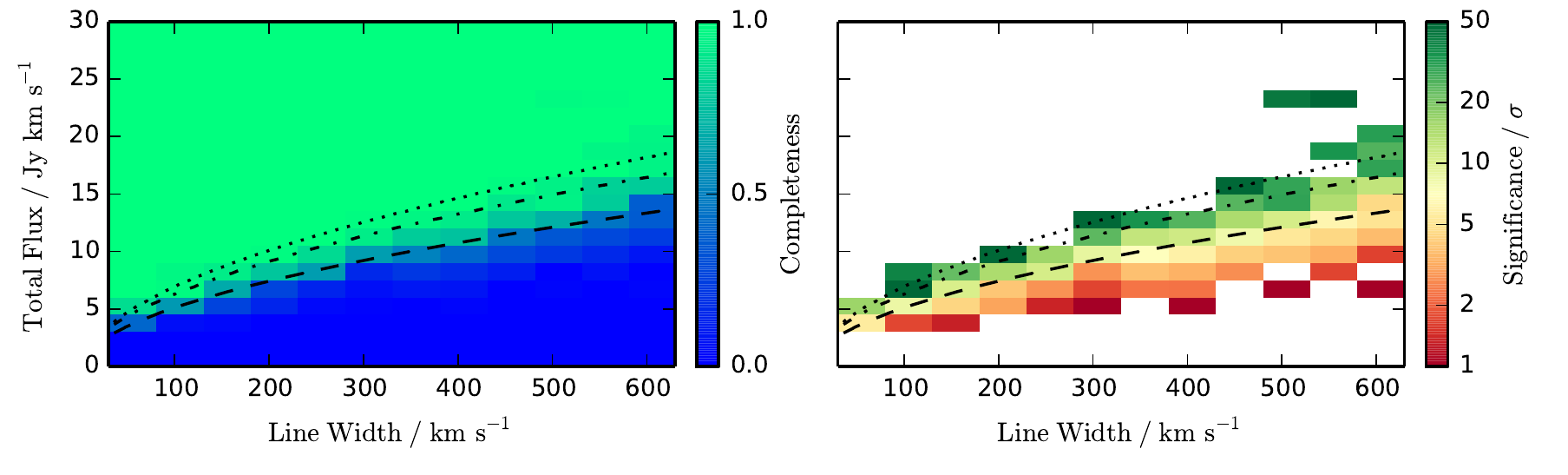}
\caption{\textit{Left panel}: Completeness as a function of linewidth and total flux. \textit{Right panel}: Significance of the derived completeness level, as determined by bootstrap resampling. Bins without an assigned significance do not show any variance. In both panels, the dashed, dash-dotted, and dotted lines indicate the 50\%, 95\%, and 99\% completeness level as determined from our completeness model.}
\label{fig:completeness_wavelet}
\end{figure*}

\subsection{Completeness}
\label{sec:completeness}

After the parametrization, we crossmatch the sources detected by the pipeline with the input catalog to evaluate the completeness. We define the completeness as the fraction of sources detected at a given flux and linewidth. A source counts as detected if there is a candidate that is less than half a beam away and whose systemic velocity lies inside the interval $v \pm w_{50}/2$ of the simulated source. As shown in Sect. \ref{sec:parameter_accuracy}, the actual positional and systemic velocity accuracy is much more precise than this matching criterion.

Figure~\ref{fig:completeness_wavelet} shows the results for the source-finding approach described in Sect. \ref{sec:sourcefinding}. We show the completeness as a function of total flux and linewidth as these are the two key source parameters obtained from single-dish \ion{H}{I} surveys. The significance of the completeness level is derived using bootstrap resampling \citep{efron1979} and dividing each completeness bin by its bootstrap error. To have an analytic description of the completeness of the survey, we fit the following model derived from a logistic function or sigmoid to the binned completeness:
\begin{equation}
C\left(F_{\rm tot}\left[\rm Jy\,km\,s^{-1}\right],w_{50}\left[\rm km\,s^{-1}\right]\right) = \left[\exp\left(-\frac{F_{\rm tot} - a_1\,w_{50}^{a_2}}{a_3\,w_{50}^{a_4}}\right) + 1\right]^{-1}\ .
\end{equation}
Here the coefficients $a_1$ and $a_2$ determine the shift of the completeness, as the profiles get wider and therefore harder to detect. The coefficients $a_3$ and $a_4$ determine the increase of the width of the completeness function as wider profiles have a higher chance of being pushed above the detection limit by random fluctuations.
Theoretically, one expects a shift of the completeness level proportional to $\sqrt{w_{50}}$, meaning $a_2 \approx 0.5$. Indeed, when we fit the model to the data, we obtain
$a_1 = 0.42\pm0.02$, $a_2 = 0.540\pm0.007$, $a_3 = 0.033\pm0.009$, and $a_4 = 0.54\pm0.04$. We plot the 50\%, 95\% and 99\% completeness levels in Fig. \ref{fig:completeness_wavelet}, respectively.

\subsection{Parameter accuracy}
\label{sec:parameter_accuracy}

To determine the accuracy of the parametrization pipeline, we compare various measured parameters with their input value. The parameters of interest are position, redshift, peak flux, linewidth, and total flux. To quantify bias and error in the parameters, we combine the pipeline output from the low and high signal-to-noise data cubes. For each parameter, we calculate the absolute error, $\Delta V = V_{\rm measured} - V_{\rm true}$, as a function of the input parameter that predominantly determines its accuracy. We split the data in ten bins of equal source counts and calculate the median and the range that contains 95\% of the data in any individual bin. Although it might be of interest to derive a probability density that relates the measured value to the true value, as in $p\left(\theta^{\rm true}\, \middle|\, \theta^{\rm measured}\right)$, this is not straight forward, as the errors exhibit strong heteroscedasticity and asymmetry. A complete statistical modeling will be the topic of a later investigation.

\subsubsection{Peak flux}
\label{sec:peak_flux}

\begin{figure}[t]
\includegraphics{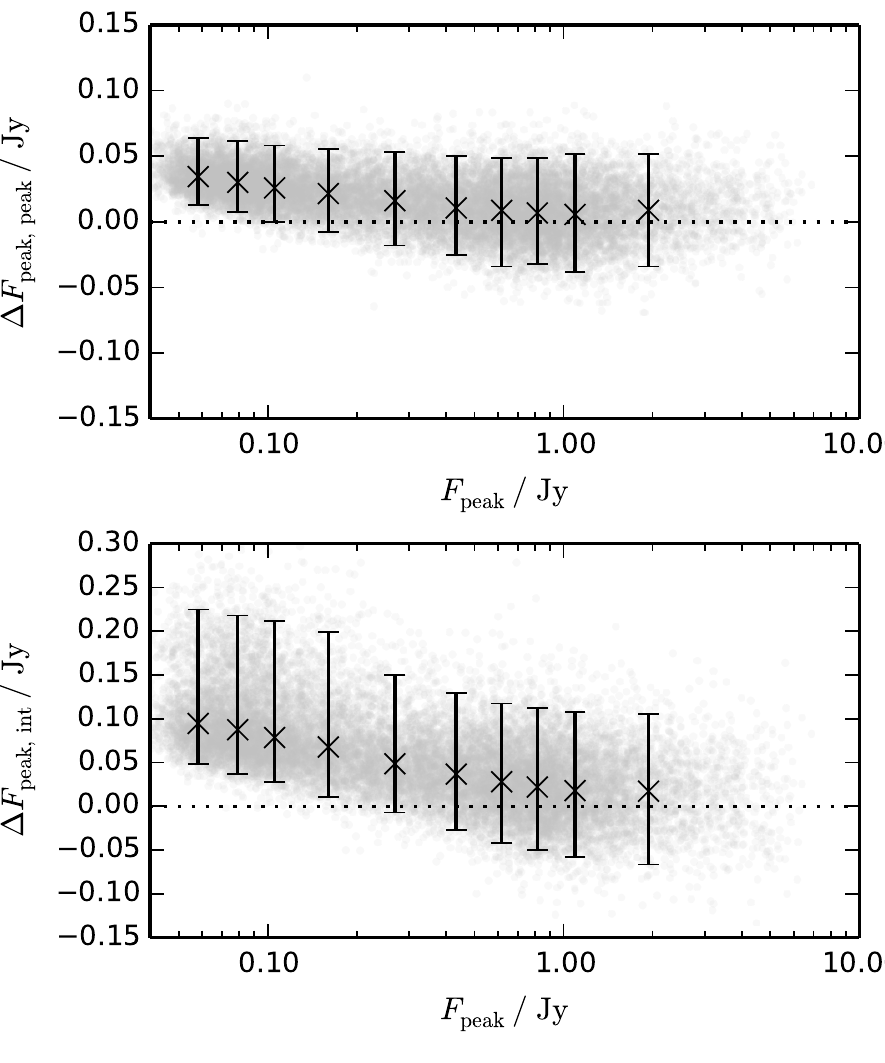}
\caption{\textit{Top panel}: Absolute error in the peak flux as a function of true peak flux as measured from the peak spectrum. The crosses and error bars indicate the median and the 95\% confidence region. The gray dots show the distribution of the individual measurements. \textit{Bottom panel}: Same as above but for the peak flux measured from the integrated spectrum.}
\label{fig:accuracy_peak_flux}
\end{figure}

The peak flux of a source profile is not a parameter of the source. This can be easily seen if we consider the same line profile at different spectral resolutions. The lower the resolution, the lower the measured peak flux, since peaks are averaged out. It is nonetheless an interesting parameter, since the peak signal-to-noise ratio determines the accuracy of the linewidth measurement and therefore also the determination of the redshift of a source. 
Assuming an unresolved source, the peak flux can be measured from the peak profile, which we reconstruct according to the formula,
\begin{equation}
\label{eq:peak_weight}
F\left(v\right) = \frac{F\left(p_x,p_y,v\right)}{B\left(p_x - c_x, p_y - c_y\right)}\ .
\end{equation}
Here, $p_x$ and $p_y$ are the position of the brightest pixel in a moment map and $c_x$ and $c_y$ are the calculated centroid coordinates of the source. The function $B(x,y)$ is the normalized beam function, which we approximate by a Gaussian with an FWHM of 10\farcm 8. This weighting accounts for the fact that the peak flux is only contained in the data if a source is located directly in the center of a pixel and needs to be extrapolated otherwise. This correction is typically about 10\%. For resolved sources, we can not measure the peak flux from the peak spectrum, and we use an integrated spectrum instead. Since this integrated spectrum is much noisier than a peak spectrum, the measured peak flux is expected to have a larger scatter.

In Fig. \ref{fig:accuracy_peak_flux}, we show the absolute error in the peak flux for both methods. In both cases, the extent of the 95\% confidence region is roughly constant down to $0.3\rm\,Jy$, but the distribution gets more asymmetric and skewed for lower peak fluxes. This can be explained by considering that there is an increasing probability that the noise fluctuations produce a larger value than the actual peak of the profile for a low peak signal-to-noise. Since we measure the peak flux from the largest value in the spectrum, this leads to a systematic overestimation of the peak flux. This also explains the larger scatter in the integrated spectrum as it is noisier than the peak spectrum.

\subsubsection{Position and redshift}

\begin{figure}[t]
\includegraphics{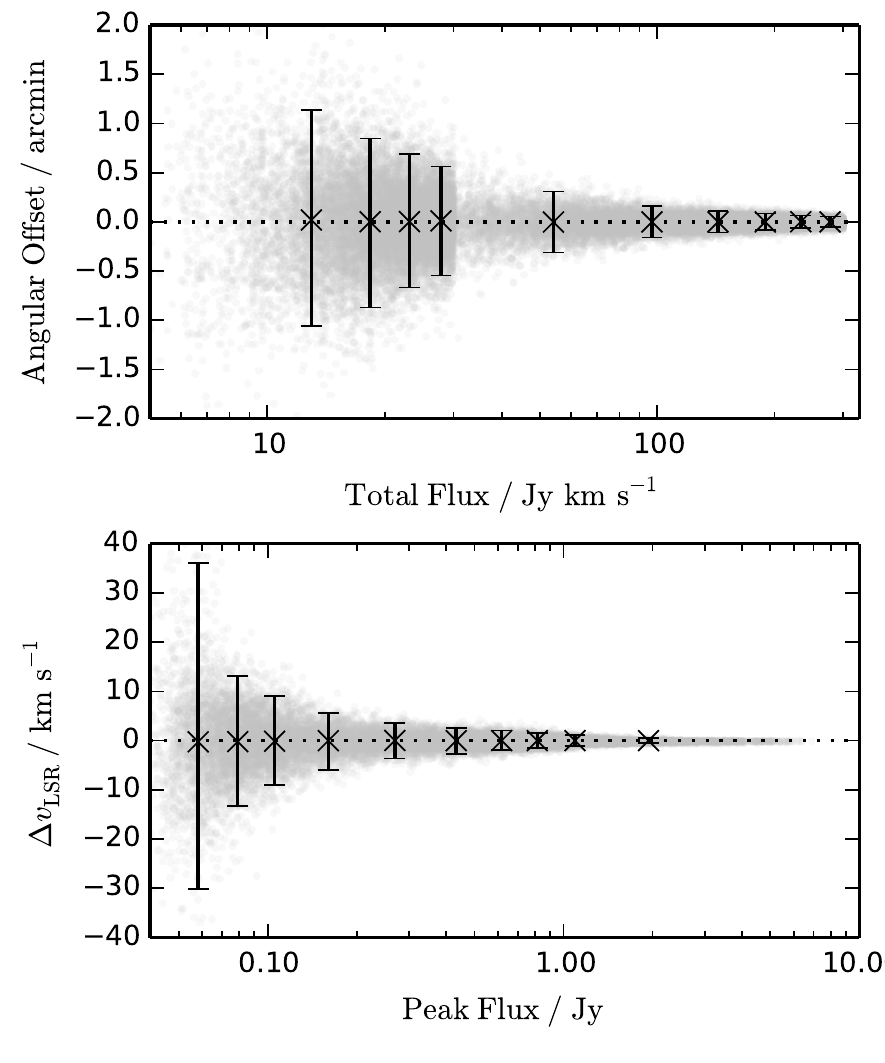}
\caption{\textit{Top panel}: Positional accuracy as a function of total flux. The crosses and error bars indicate the median and the 95\% confidence region. The gray dots show the distribution of the individual measurements. \textit{Bottom panel}: Redshift accuracy as a function of peak flux. Gray dots, crosses, and error bars have the same meaning as above.}
\label{fig:accuracy_position_redshift}
\end{figure}

Figure \ref{fig:accuracy_position_redshift} displays the accuracy for the positional parameters. As the angular position is measured from the velocity-integrated map, its accuracy depends on the total flux of the source. The angular position of a source is typically determined with a scatter of less than a tenth of the angular resolution. Since the redshift is measured as the midpoint between the width of the line, its accuracy is dependent on the peak signal-to-noise ratio. The redshift is determined with subchannel accuracy, even for fairly low signal-to-noise sources. Both parameters do not show any measurable bias.

\subsubsection{Linewidth}

\begin{figure}[t]
\includegraphics{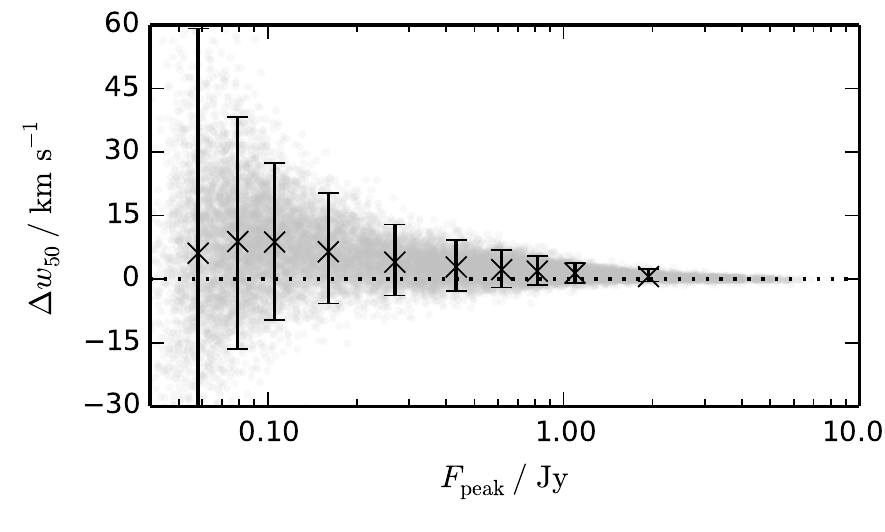}
\caption{Absolute error in $w_{50}$ as a function of peak flux. The crosses and error bars indicate the median and the 95\% confidence region. The gray dots show the distribution of the individual measurements.}
\label{fig:accuracy_linewidth}
\end{figure}

Since the measurement of the $w_{50}$ profile width relies on the peak flux, it is expected that it is similarly biased as the peak flux itself. Indeed, from Fig. \ref{fig:accuracy_linewidth}, the bias for low signal-to-noise profiles is clearly evident. Once the typical amplitude of the line profile approaches the noise level, the 50\% level of the measured peak flux is not significant enough to be easily distinguished from the noise level. Since we are using a width-maximization algorithm, the linewidth is predominantly overestimated.

\subsubsection{Total flux}

\begin{figure}[t]
\includegraphics{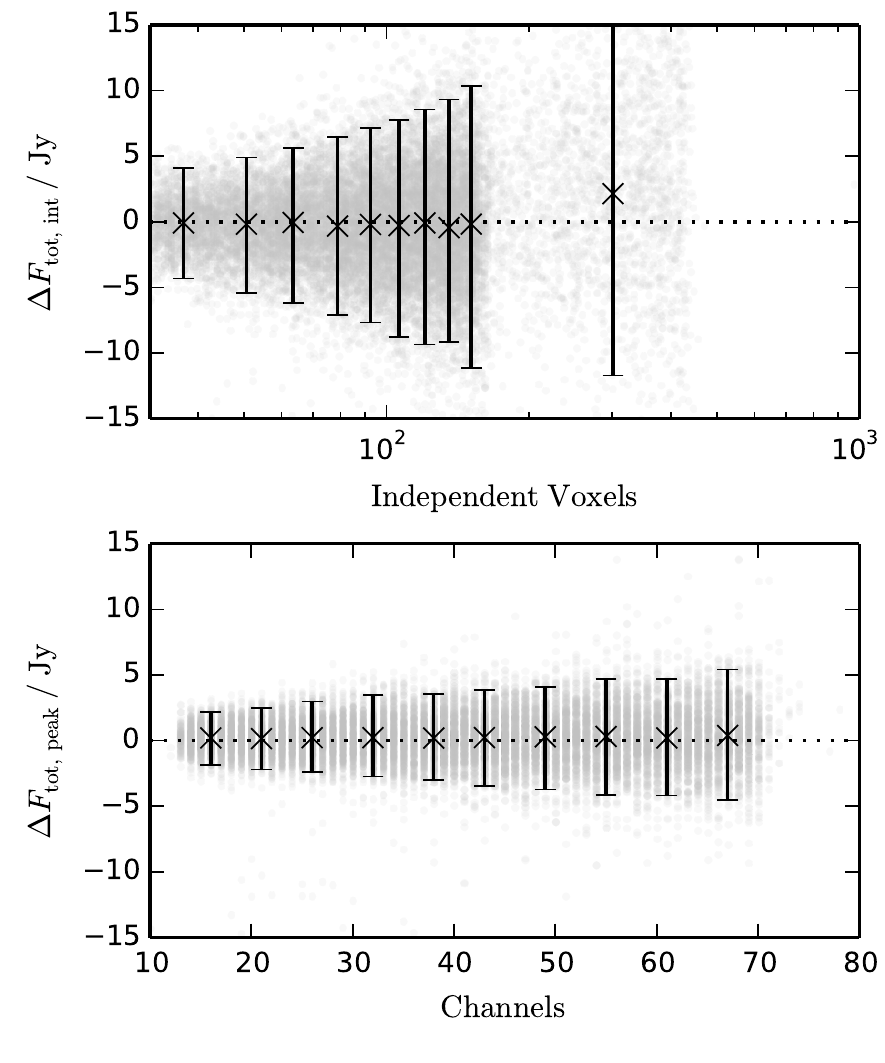}
\caption{\textit{Top panel}: Absolute error in the total flux as a function of the number of independent voxels as measured from the integrated spectrum. The crosses and error bars indicate the median and the 95\% confidence region. The gray dots show the distribution of the individual measurements. \mbox{\textit{Bottom panel}}: Absolute error in the total flux as a function of the number of independent voxels as measured from the peak spectrum. Gray dots, crosses, and error bars have the same meaning as above.}
\label{fig:accuracy_total_flux}
\end{figure}

Just like the peak flux, the total flux can be measured in two ways: In the case of an unresolved source, one can simply sum up the corrected peak profile. In this case, the accuracy of the total flux is primarily determined by its linewidth. If the source is resolved, one has to use the integrated spectrum, which again has a higher noise level. In both cases, it is expected that the uncertainty in the total flux is proportional to the square root of statistically independent values summed up. In the case of the peak spectrum, this is the number of channels, whereas in the case of the integrated spectrum the number of statistically independent voxels.

In Fig. \ref{fig:accuracy_total_flux}, we plot the results for both cases. Since the peak spectrum is less noisy and we are simulating unresolved sources only, the scatter is very much reduced as opposed to the integrated spectrum method. Furthermore, the bin corresponding to the highest amount of independent voxels (three-dimensional pixels in our data cubes) shows a clear bias. These are the cases where the aperture optimization fails due to the source having a very low signal-to-noise ratio and the mask growing as long as it adds up positive noise clusters.

\subsection{Comparison with HIPASS}

To quantify how our pipeline performs compared to manual parametrization, we take the error estimates from the \mbox{HIPASS} survey as derived by \citet[][hereafter Z04]{2004MNRAS.350.1210Z}.
 A major difference between the error estimates in Z04 and our estimates is that the errors in two parameters of interest, total flux and linewidth, are either given as a function of themselves or as a constant value for all sources. In the previous section, we have shown that the errors in these parameters have a clear dependence on other parameters. For this reason, we can only compare average errors.

Furthermore, the synthetic sources used by Z04 to investigate the errors on the parameters were drawn from a uniform distribution in peak flux and linewidth. This assigns sources with higher linewidth a higher total flux, which leads to a lack of faint, high-linewidth sources. These sources are particularly difficult to detect and parametrize. This also decreases their derived completeness level as they observe a counter-intuitive incompleteness for narrow sources (Z04, their Fig. 2). This is substantiated by their derived completeness from narrowband follow-up observations, as it does not exhibit reduced completeness for narrow linewidths (Z04, their Fig. 7).

A striking difference between the errors derived by Z04 and our investigation is the shape of the 95\% confidence region for the peak flux. In Sect. \ref{sec:peak_flux}, we argue that it is expected for the peak flux of the profile to be a biased measurement. The investigations of Z04 do not show a varying bias, and they adopt a constant error, which corresponds to a 95\% confidence region of $\pm 22\rm\,mJy$, which is slightly less than twice the typical noise in their spectra. We also observe that the span of our 95\% confidence region is approximately twice the typical noise in our spectra. The absence of a strong, positive bias toward low signal-to-noise ratios in Z04 indicates at a more complicated measurement method, which is not explained by the authors.

For the $w_{50}$ linewidth, Z04 adopt a $1\sigma$ error of $7.5\rm\,km\,s^{-1}$, which corresponds to a 95\% confidence region of $\pm 15\rm\,km\,s^{-1}$. The 95\% confidence region derived for our pipeline reaches this accuracy at a peak signal-to-noise ratio of approximately 10. Below that, the error grows significantly but turns out to be much smaller for higher signal-to-noise ratios. However, Z04 note that the histogram of roughly one-third of the synthetic sources used to measure this error are better fit by a Gaussian with a dispersion of $25\rm\,km\,s^{-1}$, which in turn corresponds to a 95\% confidence region of $\pm 50\rm\,km\,s^{-1}$. Since the errors derived from our simulations are highly asymmetric for lower signal-to-noise ratios, it is not straight forward to compare the accuracy of both surveys in this regime. However, we note that only for the faintest sources does our 95\% confidence region span $100\rm\,km\,s^{-1}$.

For the total flux, Z04 calculates the $1\sigma$ error according to $\sigma_{F_{\rm tot}} = 0.5 \sqrt{F_{\rm tot}}$, which is equivalent to a 95\% confidence region of $\pm 3\rm\,Jy\,km\,s^{-1}$ at their 99\% completeness level. This matches very well the mean confidence region derived for our measurement of $F_{\rm tot}$ from the peak spectrum. Our pipeline is even more accurate for sources with an extent of less than 40 spectral channels. As this already corresponds to over $400\rm\,km\,s^{-1}$ and most galaxies detected in shallow \ion{H}{I} surveys seem to have a $w_{50}$ of less than $200\rm\,km\,s^{-1}$ \citep{Zwaan21042010}, the mean error for unresolved galaxies in EBHIS is expected to be smaller than in HIPASS.

The errors in redshift and angular position given by Z04 correspond to a 95\% confidence region of $\pm 12.8\rm\,km\,s^{-1}$ and $\pm 1.64\rm\,arcmin$, respectively. The 95\% confidence regions for the respective parameters are smaller for the whole range of sources detected in our simulations with the exception of the redshift for very faint sources. This highly increased accuracy is certainly caused by the higher angular and spectral resolution of EBHIS data but also shows that the employed algorithms show very good performance.

Overall, even though EBHIS does not have HIPASS sensitivity, the parameter accuracy achieved by our fully automated pipeline seems to be on par and partly even surpassing the error estimates for HIPASS.

\section{Classification}
\label{sec:classification}

\begin{figure*}
\centering
\includegraphics{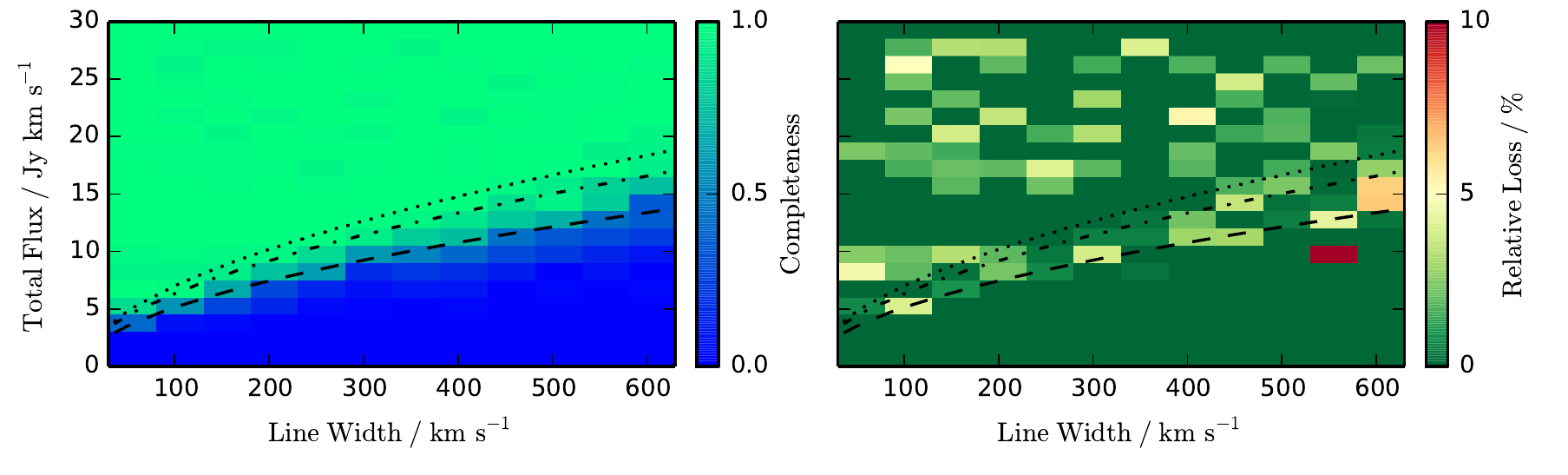}
\caption{\textit{Left}: Completeness after classification as a function of linewidth and total flux. The dashed, dash-dotted, and dotted lines indicate the 50\%, 95\%, and 99\% completeness level as determined from our completeness model. \textit{Right}: Relative loss in completeness as compared to Fig. \ref{fig:completeness_wavelet}. As a guide, the lines show the same model as shown in the left panel.}
\label{fig:classified_completeness}
\end{figure*}

Although state-of-the-art \ion{H}{I} surveys employ various levels of automation through the use of source finders or interactive parametrization, the decision whether a source constitutes a real detection is still left for an astronomer. Apart from the tedious nature of looking at thousands of false positives, this approach comes with another pitfall, especially when it comes to marginal detections: Different astronomers might disagree on an individual source and even the performance of an individual astronomer can differ from day-to-day. For example, \cite{2004MNRAS.350.1195M} minimize the risk of misidentification by having three individual astronomers look at more than 140\,000 candidates to extract 5\,000 true detections.

For the data volume expected from SKA and its pathfinder experiments, this approach is not feasible anymore. Even assuming that we can perfectly distinguish sources from the noise, the amount of false positives scales with data volume, as most of them are produced by defects in the data, such as radio-frequency interference (RFI). Here, automated decisions about the true nature of a source have to be made from the measured parameters alone by employing machine learning algorithms. 

\cite{doi:10.1142/S0218271810017160} present a good overview of machine learning in astronomy and highlight a large range of examples of how different machine learning algorithms have been applied to astronomical data sets. We decide to implement classification for EBHIS by using an artificial neural network (ANN).

\subsection{Neural network implementation and training}

Among the advantages of ANNs is their ability to take a large number of input parameters to come to a classification conclusion. They are robust to redundant parameters in the input, which makes them especially useful when it is not known which parameters contain the most information. This alleviates the need to perform a careful parameter selection beforehand by either manually excluding parameters or pursuing a more sophisticated approach like principal component analysis \citep{doi:10.1142/S0218271810017160}. A considerable drawback of ANNs is their large number of free parameters that need to be trained. To obtain useful predictive results from an ANN, a good training data set has to be created. This training set has to cover the whole parameter range of interest and needs to be sufficiently large.

The simplest fully connected ANN with no hidden layers can be expressed by
\begin{equation}
O(x) = A\left(b + W x\right)\ .
\end{equation}
Here, $O(x)$ is the output vector, $A(x)$ is the activation function, $b$ is the bias vector, $W$ is the weight matrix, and $x$ is the input vector. Complex networks are built by replacing $x$ with the output of another simple network. In our networks, all activation functions are hyperbolic tangents with the exception of the last one. Here, we use the softmax function whose $i$th output value is given by $\text{softmax}_i(x) = e^{x_i}\left(\sum_j e^{x_j}\right)^{-1}$. Using this function the sum of the output layer will always equal unity. Each output node represents one class, and we assign candidates to the class corresponding to the node with the highest output. The ANN is implemented using the Python package Theano \citep{bergstra+al:2010-scipy}, which offers a number of tools to implement neural networks using linear algebra using the above mentioned formalism.

In preparation for the training, we compile a training data set containing equal parts of true and false positives. As the bulk of false positives is caused by effects, which are not included in our simulations, we do not have a sufficient number of false positives for ANN training. Instead, we manually classify the pipeline output from EBHIS data cubes that show a larger than average number of defects of different origins, which are mostly residual RFI and unstable baselines from bright continuum sources. The resulting list of candidates contains 7\,557 entries 52 of which are actual galaxies that we identify by eye and cross-check with existing redshift catalogs. To obtain a balanced training data set, we add detected sources from our simulation. The final training data set contains 14\,382 entries.

For each entry in the training data set, we compile a feature vector from the values measured by the pipeline. From both the sum and peak spectra, we include in the feature vector:
\begin{itemize}
\item total and peak flux
\item integrated and peak signal-to-noise ratio
\item $w_{50}$ linewidth
\item skewness and kurtosis of the profile
\end{itemize}
From the velocity-integrated maps of each source, we include
\begin{itemize}
\item major and minor axis length and their ratio and
\item differential and cumulative surface brightness profile.
\end{itemize}

Once all feature vectors are compiled, we whiten the scatter of the individual parameters by subtracting their mean and dividing by their standard deviation to ensure a good training behavior \citep{lecun1998efficient}. These scaling parameters are stored, since any catalog that is to be classified by an ANN trained with the training data set needs to be shifted and scaled by the same values.

Since there are no good ad-hoc rules of how to choose the hyper-parameters of an ANN, as in the number of layers and nodes per layer, we perform a grid search to find the optimal ANN for our classification task. The different networks are trained using the back-propagation algorithm with stochastic gradient descent \citep{Robbins:1951aa, rumelhart1986learning, bottou-tricks-2012}. To avoid over-fitting to the training data, we train the networks on a random subsample of 70\% of the training data and use the remaining 30\% as a validation data set. The purpose of the validation data set is to stop the training of the ANN once the classification error only decreases on the training data but increases on the validation data. This strategy, called early stopping, stops the training of the ANN once it no longer learns general properties of the data but instead learns the properties of the training data set. We furthermore employ L2 regularization to avoid saturated weights \citep{2012arXiv1206.5533B}. 

In our grid search, we vary the number of nodes per layer, the number of layers, the learning rate $\eta$ and the L2 regularization parameter $\lambda$. We find that an ANN with one hidden layer, which is two layers of weights and biases, is sufficient for our classification task. This means that the two classes in our problem, which are true and false positives, are not efficiently linearly separable in the parameter space created by our 38-element feature vector \citep{haykin13neural}. Furthermore, our grid search prefers small L2 regularization ($\lambda = 1\times 10^{-6}$)
and a fast learning rate  ($\eta = 0.1$). The percentage of mis-classified sources is commonly below $1\%$ and reaches $0.6\%$ for the best case. Based on these results, we choose an ANN with 40 nodes in the hidden layer, giving it the architecture 38 -- 40 -- 2.

\subsection{Performance}

Since the classification is not 100\% accurate, it is obvious that the usage of an ANN has an impact on the completeness and reliability of the survey. We investigate the impact on completeness by classifying all detected sources from our simulated data cubes. In Fig.~\ref{fig:classified_completeness}, we show the survey completeness after classification and the relative decrease in completeness, as compared to Fig.~\ref{fig:completeness_wavelet}. On average, we lose 0.6\% of the detected sources due to misclassification. We again fit the completeness, as described in Sect.~\ref{sec:completeness}, and obtain the parameters 
$a_1 = 0.41\pm0.03$, $a_2 = 0.54\pm0.01$, $a_3 = 0.03\pm0.01$, and $a_4 = 0.58\pm0.07$.
Since the loss in sources is nearly uniform over the $w_{50}$ - $F_{\rm tot}$ plane, the shape of the transition from 0\% to 100\% completeness does not change noticeably. 

\begin{figure}[t]
\centering
\includegraphics{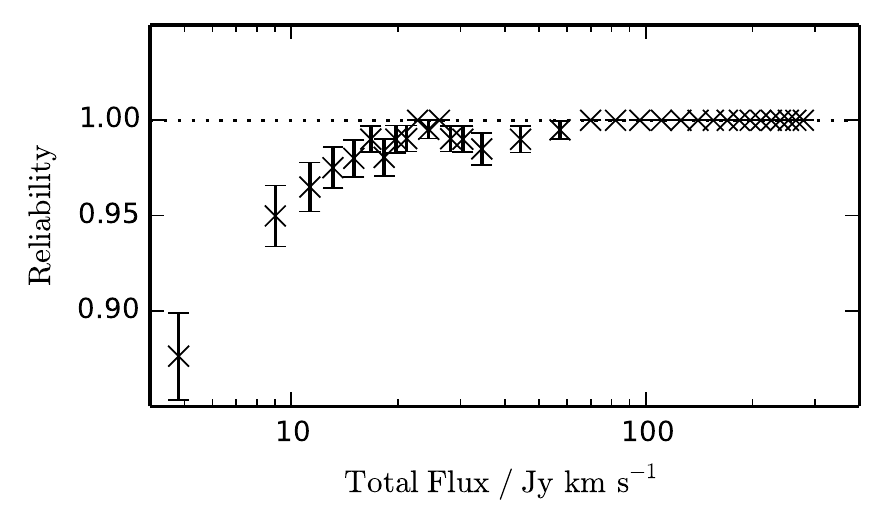}
\caption{Classification reliability as a function of measured total flux. The bins have been chosen to have 200 sources each. The error bars are estimated using bootstrap resampling.}
\label{fig:reliability}
\end{figure}

Another measure of the performance of the ANN is the reliability, which is the probability that a source that is classified as true is actually a real source. In Fig.~\ref{fig:reliability}, we plot the reliability as a function of measured flux. Note that this is different from the plots in the previous sections and the completeness plots, as there is no real flux for false positives. The reliability starts at a very high level of over $0.9$ overall and rises to $1.0$, as the total flux approaches the survey completeness limit. The high reliability is a result of the accurate classification and careful noise-modeling during the source finding stage. It has been shown previously that the 2D-1D wavelet denoising approach is very unlikely to generate false positives by picking up noise clusters and exhibits a high reliability \citep{2012PASA...29..318P}. The small drops in reliability between $30\rm\ Jy\,km\,s^{-1}$ and $100\rm\ Jy\,km\,s^{-1}$ can be attributed to broadband RFI events, which are bright and closely mimic the shape of point sources \citep{2010rfim.workE..42F}. The mean reliability for the test data set is 99.7\%. As a comparison, \cite{2004MNRAS.350.1210Z} determined an average reliability for HIPASS of 95\%, which includes improved reliability through follow up observations.

The performance of the ANN shown here is derived mostly from simulated sources and true artifacts in the data. It is expected that the true performance with real data will vary. In particular, the range of parameters spanned by our simulated sources is limited in comparison to the range spanned by the false positives. We will therefore use the ANN trained on simulated sources to extract a first set of true sources from EBHIS and re-train the ANN using real data only. The high reliability of this classification approach should make it feasible to quickly generate a set containing equal parts true and false positives.

\section{Conclusions}
\label{sec:conclusions}

In this paper, we present our implementation of a fully automated source finding, parametrization, and classification pipeline for the Effelsberg-Bonn HI Survey, EBHIS. We conclude the following:
\begin{enumerate}
\item Because of its automated nature, we can test the pipeline with simulated data cubes containing a total of 24\,000 sources. With this data set, we derive precise confidence regions for the errors in completeness, parametrization, and classification.
\item Wavelet denoising is a powerful tool for source finding for \ion{H}{I} surveys. The derived completeness shows a smooth transition from 0\% to 100\%, and no bright sources are missed. The proven robustness of the denoising scheme against common data defects makes it plausible that it will fare similarly well on real data.
\item Our automated algorithms enable unsupervised parametrization. Using simulated data sets, we show that our unsupervised pipeline is competitive with the accuracy achieved in the manually parametrized and more sensitive HIPASS survey. We derive 95\% confidence regions for the main parameters of interest: position on the sky, redshift, linewidth, and total flux. The position on the sky is typically determined to less than an arc minute precision. Except for the faintest sources, we determine the redshift to sub-channel accuracy, which is less than $10.24\rm\,km\,s^{-1}$. Using a bilateral filter, we decrease the bias and scatter observed when measuring the width of the line profile at the 50\% level of its peak flux. For a peak signal-to-noise ratio of five, we determine the linewidth with less than $20\rm\,km\,s^{-1}$ error. For unresolved sources, we measure the total flux with less than $5\rm\,Jy\,km\,s^{-1}$ error. 
\item To automate the task of classification, we train an artificial neural network to discern false positives from real sources. We only lose 0.6\% of detected sources due to misclassification and achieve an average reliability of 99.7\%. The lost sources do not affect the shape of the transition from 0\% to 100\% completeness. The high reliability makes it possible to use the current pipeline to compile a new training data set from real data only. An ANN trained on this way should perform even better on real data.
\item Our results show that completely unsupervised source extraction is a feasible and competitive approach for large-scale \ion{H}{I} surveys. The developed pipeline is applicable to any single-dish \ion{H}{I} survey comparable to EBHIS, such as HIPASS or ALFALFA. With a more refined approach to resolved sources, the pipeline should also be able to serve as a first source extraction step for future \ion{H}{I} surveys with ASAKP and WSRT/Apertif. 
\end{enumerate}

\begin{acknowledgements}
The authors thank the Deutsche Forschungsgemeinschaft (DFG) for support under grant numbers KE757/7-1, KE757/7-2, KE757/7-3 and KE757/9-1.
The authors would like to thank the anonymous referee for his valuable comments and the language editor for his careful corrections.
L.F. is a member of the International Max Planck Research School (IMPRS) for Astronomy and Astrophysics at the Universities of Bonn and Cologne.
This research has made use of NASA's Astrophysics Data System.
This research has made use of the NASA/IPAC Extragalactic Database (NED) which is operated by the Jet Propulsion Laboratory, California Institute of Technology, under contract with the National Aeronautics and Space Administration. 
This research has made use of the SIMBAD database, operated at CDS, Strasbourg, France.
Some results are based on observations with the 100-m telescope of the Max-Planck-Institut für Radioastronomie (MPIfR) at Effelsberg.
We acknowledge the use of NASA's \textit{SkyView} (http://skyview.gsfc.nasa.gov) located at NASA Goddard Space Flight Center.
Some figures have been prepared using the Kapteyn Package \citep{KapteynPackage}
\end{acknowledgements}

\bibliography{ebhis_sfpc}

\end{document}